# Multispectral Quantitative Phase Imaging Using a Diffractive Optical Network

Che-Yung Shen[1,2,3†], Jingxi Li[1,2,3†], Deniz Mengu[1,2,3], and Aydogan Ozcan[1,2,3*]

[1]Electrical and Computer Engineering Department, University of California, Los Angeles, CA, 90095, USA

[2]Bioengineering Department, University of California, Los Angeles, CA, 90095, USA

[3]California NanoSystems Institute (CNSI), University of California, Los Angeles, CA, 90095, USA

[†]These authors contributed equally to the work

[*]Correspondence to: ozcan@ucla.edu

**Abstract**

As a label-free imaging technique, quantitative phase imaging (QPI) provides optical path length information of transparent specimens for various applications in biology, materials science, and engineering. Multispectral QPI measures quantitative phase information across multiple spectral bands, permitting the examination of wavelength-specific phase and dispersion characteristics of samples. Here, we present the design of a diffractive processor that can all-optically perform multispectral quantitative phase imaging of transparent phase-only objects in a snapshot. Our design utilizes spatially engineered diffractive layers, optimized through deep learning, to encode the phase profile of the input object at a predetermined set of wavelengths into spatial intensity variations at the output plane, allowing multispectral QPI using a monochrome focal plane array. Through numerical simulations, we demonstrate diffractive multispectral processors to simultaneously perform quantitative phase imaging at 9 and 16 target spectral bands in the visible spectrum. These diffractive multispectral processors maintain uniform performance across all the wavelength channels, revealing a decent QPI performance at each target wavelength. The generalization of these diffractive processor designs is validated through numerical tests on unseen objects, including thin Pap smear images. Due to its all-optical processing capability using passive dielectric diffractive materials, this diffractive multispectral QPI processor offers a compact and power-efficient solution for high-throughput quantitative phase microscopy and spectroscopy. This framework can operate at different parts of the electromagnetic spectrum and be used for a wide range of phase imaging and sensing applications.

# 1 INTRODUCTION

Quantitative phase imaging (QPI) is a label-free imaging technique that measures the optical path length variations induced by a sample[1–3]. By evaluating the phase shifts that light undergoes when interacting with a specimen, QPI derives information about the refractive index and thickness of the sample. Numerous QPI methodologies have been reported in the literature, including Fourier phase microscopy[4], Hilbert phase microscopy[5], digital holographic microscopy[6–12], and spatial light interference microscopy[13], among others. QPI serves as a versatile tool in biomedical fields such as cell biology[14,15], pathology[16–18], and biophysics[19], covering various applications including real-time cell growth and behavior[20,21], cancer detection[22,23], pathogen sensing[24,25], and the study of subcellular structures and processes[26]. Additionally, QPI has also found applications in other fields, including materials science and nanotechnology, such as the characterization of thin films, nanoparticles, and fibrous materials, elucidating their unique optical and physical properties[27–29].

A typical QPI device is composed of two primary components: (1) an imaging front-end responsible for conducting optical interferometry to convert the desired phase information into intensity signals that can be captured using a digital image sensor, and (2) a digital processing back-end tasked to perform the essential image processing and reconstruction of quantitative phase images based on these signals. Recently, propelled by advancements in deep learning, the versatility and performance of digital backends used in QPI devices have been extensively enhanced. For instance, by leveraging the massively parallel computing power of graphics processing units (GPUs), the image reconstruction speed and throughput of QPI devices have seen substantial improvements[30–32]. Moreover, deep learning algorithms can be utilized to interpret QPI images to accomplish advanced tasks such as the classification and detection of specific target objects[33–35], while also assisting in solving inverse problems in QPI, which include phase retrieval[31,32,36–41], aberration correction[42,43], depth-of-field extension[44,45] and image modality transformations[18,46].

By and large, QPI devices that are commonly employed utilize a monochromatic source, thereby providing phase information corresponding to a single wavelength. Building upon this, multispectral QPI systems[47–51] have also been developed to simultaneously acquire a sample's phase information across multiple spectral bands. These advancements facilitated the detection of samples' dispersion signatures and their spatial distribution, finding use in diverse areas such as quantifying hemoglobin concentration in erythrocytes[47,51] and mapping the spatial refractive index distribution of the retina[52]. However, these existing multispectral QPI techniques require bandpass filters (e.g., color filter wheels) or a tunable laser source to obtain the desired QPI signals at individual wavelengths. This can introduce considerable cost and complexity to the QPI system hardware, enlarge its footprint, and potentially result in chromatic aberrations in the imaging system if left uncorrected. Moreover, the digital processing of the resulting multispectral holographic images further adds to the computational load of the digital back-end.

In this work, we introduce a multispectral QPI system designed based on a broadband diffractive optical neural network, which can simultaneously obtain the quantitative phase profiles of dispersive phase objects across a wide range of spectral bands within a single snapshot. This diffractive multispectral QPI processor comprises a series of spatially-engineered dielectric diffractive layers, collectively forming an optical neural network[53–65], as illustrated in **Figure 1**. After its optimization through deep learning, this optical network transforms the multispectral phase information of input objects into a distinct intensity distribution at the output field-of-view (FOV) that spatially encodes the object phase information corresponding to each target spectral band separately. This intensity distribution can be measured using a monochrome image sensor array, and, following a standard demosaicing and normalization process, results in quantitative phase images of the input object at the target spectral bands of interest. Our diffractive QPI processor design has a very compact footprint that axially spans only ~$100\lambda_m$, where $\lambda_m$ represents the average wavelength of the target spectral band. It incorporates passive optical elements, and can instantaneously perform end-to-end multispectral QPI directly on an input object or scene, offering a powerful alternative to traditional digital QPI frameworks that depend on specialized hardware for the optical relay, spectral filtering and digital phase recovery steps.

To numerically demonstrate the feasibility of our concept, we trained two diffractive designs that can perform multispectral QPI at 9 and 16 unique spectral bands, evenly distributed within the visible spectrum from 450 to 700 nm. Both of these designs utilize 10 successive diffractive layers that are laterally engineered at a feature size of ~225 nm, forming compact diffractive systems axially spanning ~64.8 μm. The performance of these diffractive multispectral QPI designs was tested based on numerical simulations, demonstrating average peak signal-to-noise ratio (PSNR) values of >17.0 dB and >16.6 dB and structural similarity index measure (SSIM) values of >0.77 and >0.72 for the 9- and 16-channel designs, respectively. Our cross-talk analyses also indicate that, for a given group of sensor pixels designated to a target spectral band, the optical power received from the target spectral band was, on average, more than 7× higher than the mean power received from the other wavelengths (representing cross-talk), showcasing the success of our spectral filtering performance. We also quantified the spatial resolution of our QPI designs to reveal that our 9- and 16-channel diffractive designs could resolve phase features with a linewidth of ≥5.4 μm and ≥7.2 μm, respectively. The external generalization performance of our diffractive multispectral QPI designs was further validated through blind tests using images of Pap smear samples, showcasing its versatility to accommodate a wide range of distinct spatial features never seen in the training of the diffractive networks.

Our diffractive multispectral QPI design is also scalable according to operating wavelengths. While we employed the visible spectrum for the numerical testing of our diffractive multispectral QPI designs, they can be adapted for operation at other wavelengths by simply scaling their dimensions proportional to the new spectral band of operation; this feature offers particular value for different parts of the electromagnetic spectrum where high-density, large-area spectral filter arrays are not widely available or are cost-prohibitive. Although we employed the visible spectrum for the numerical testing of our diffractive multispectral QPI designs, their intrinsic scalability also allows for adaptation to the other parts of the electromagnetic spectrum, especially those where high-density and large-area spectral filter arrays are not widely available or are cost-prohibitive. Moreover, since the diffractive layers in our designs use isotropic dielectric materials, their quantitative phase imaging operation is insensitive to the input polarization state of the illumination light. By amalgamating these unique advantages, our diffractive multispectral QPI framework offers transformative opportunities for designing quantitative phase microscopy and spectroscopy systems. These diffractive systems can be integrated with monochrome optoelectronic sensor arrays or focal plane arrays at different parts of the spectrum, resulting in high-throughput on-chip imaging and sensing devices that can find various applications in e.g., biomedical imaging, material science, and environmental monitoring.

## 2 RESULTS

Figure 1 illustrates a schematic of our diffractive multispectral QPI framework. As shown in Figure 1a, a dispersive transparent sample is illuminated by broadband spatially-coherent light at the input plane of the system. This broadband illumination can be viewed as a set of plane waves at distinct wavelengths $\{\lambda_1, \lambda_2, ..., \lambda_{N_w}\}$, organized in order from the longest to the shortest wavelength. Situated after the input plane, the broadband diffractive network comprises multiple modulation layers made of dielectric materials, where each diffractive layer is coded with the same number of spatially engineered diffractive features that have a lateral size of $\sim \lambda_{N_w}/2$ and a trainable/learnable thickness, providing a phase modulation range covering $0\text{-}2\pi$ for all the illumination wavelengths. These diffractive layers are connected to each other and the input/output planes through light diffraction in free space (air).

As outlined in Figure 1b, the optical fields $\{\boldsymbol{i}_w\}$ ($w \in \{1,2,...,N_w\}$) immediately following a phase-only transmissive object exhibit unique phase profiles $\{\boldsymbol{\Psi}_w\}$ at each transmitted wavelength $\lambda_w$, and can be represented as $\{\boldsymbol{i}_w = e^{j\boldsymbol{\Psi}_w}\}$ ($\boldsymbol{\Psi}_w \in \mathbb{R}^{N_x^{(\Psi)} \times N_y^{(\Psi)}}$). These complex fields at different wavelengths are simultaneously processed by the diffractive deep neural network (D²NN) to produce output fields $\{\boldsymbol{o}_w\}$, i.e., $\boldsymbol{o}_w = \text{D}^2\text{NN}\{e^{j\boldsymbol{\Psi}_w}\}$. The intensity profiles of these outputs are subsequently sampled by a monochrome image sensor that integrates all the wavelengths, resulting in an output optical intensity measurement $\boldsymbol{D}$ that can be expressed as:

$$\boldsymbol{D} = \sum_{w=1}^{N_w} \beta_w |\boldsymbol{o}_w|^2 \qquad (1).$$

Without loss of generality, we assume $\beta_w=1$ for all the wavelengths of interest. Considering the fact that the output optical intensity $\boldsymbol{D}$ measured by the image sensor is generally coupled to the illumination power and the output diffraction efficiency, the values of $\boldsymbol{D}$ cannot be immediately used to represent the quantitative phase values of the input object. To provide QPI performance that is independent of such power-related output fluctuations, we employed a simple normalization scheme, where we spatially divided the output measurements ($\boldsymbol{D}$) into an output signal region $\mathcal{S}$ and a reference signal region $\mathcal{R}$ to obtain the final quantitative phase measurements by calculating the relative ratio of the signals acquired within $\mathcal{S}$ and $\mathcal{R}$. As illustrated in Figure S1, Supporting Information, $\mathcal{R}$ is defined as a one-pixel wide frame located at the edges of $\boldsymbol{D}$, and can be further subdivided into $N_w$ sub-regions, each denoted as $\mathcal{R}_s$ ($s \in \{1,2,...,N_w\}$). Each $\mathcal{R}_s$ creates a wavelength-specific reference signal $Ref_s$, i.e.,

$$Ref_s = \frac{1}{N^{(\mathcal{R}_s)}} \sum_{(x,y) \in \mathcal{R}_s} D(x,y) \tag{2}$$

where $N^{(\mathcal{R}_s)}$ denotes the total number of image sensor pixels located within $\mathcal{R}_s$.

The obtained pixel intensities within $\mathcal{S}$, i.e., $\boldsymbol{D}_\mathcal{S}$, are spatially divided into $N_x^{(\Psi)} \times N_y^{(\Psi)}$ blocks, where each repeating block contains $\sqrt{N_w} \times \sqrt{N_w}$ pixels individually assigned to the $N_w$ target spectral bands according to the pixel-wavelength mapping relationship illustrated in Figure 1b. Accordingly, a spatial demosaicing operation $\delta[\boldsymbol{D}_\mathcal{S}, \mathcal{S}_s]$ is used to extract these pixel intensities in an interleaved manner, resulting in $N_w$ quantitative phase images, each corresponding to a unique spectral band $\lambda_s$, $s \in \{1,2,...,N_w\}$; see the Methods section for details. This demosaicing operation is similar to extracting the color image components from a raw image obtained with a Bayer filter-based image sensor. Finally, after being normalized using the reference signals $Ref_s$, the quantitative phase image $\boldsymbol{\Phi}_s$ of the diffractive processor can be obtained:

$$\boldsymbol{\Phi}_s = \frac{\delta[\boldsymbol{D}_\mathcal{S}, \mathcal{S}_s]}{Ref_s} \tag{3}$$

If our diffractive multispectral QPI network training is successful, we will have:

$$\boldsymbol{\Phi}_s \approx \boldsymbol{\Psi}_w \text{ (iff. } s = w\text{)} \tag{4}$$

Stated differently, the role of our diffractive multispectral QPI processor is to route the phase information of an input object at $N_w$ unique wavelengths into a periodically repeating virtual array of $\sqrt{N_w} \times \sqrt{N_w}$ pixels, each covering one wavelength of interest. This multispectral QPI processor forms an all-optical transformer that simultaneously performs (1) spectral-to-spatial and (2) phase-to-intensity transformations, serving the functions of both multispectral filtering and quantitative phase retrieval. More details about this processing pipeline and related mathematical expressions are provided in Figure 1b and the Methods section.

For the training of our diffractive multispectral QPI processor, we prepared a training image dataset composed of 110,000 images containing 55,000 handwritten images and 55,000 custom-created grating/fringe-like patterns. These structural patterns, including but not limited to gratings, patches, and circles, were created in an earlier work of our group[66], designed to augment the training data. During the training process, these images are randomly selected and encoded into the phase channels $\boldsymbol{\Psi}_w$ of the multispectral input fields using a dynamic range of $[0, \alpha_{tr}\pi]$. Considering the predominantly binary nature of the training images, we set the phase contrast parameter ($\alpha_{tr}$) of each input training image to be randomly sampled within 0.1 to 1, i.e., $U[0.1,1]$, thus ensuring wide coverage of different phase contrast values from $\alpha_{tr,min} = 0.1$ to $\alpha_{tr,max} = 1$. Error-backpropagation and stochastic gradient descent were used to optimize the thickness profiles of the diffractive layers, with the objective of minimizing a custom loss function $\mathcal{L}$ defined based on the mean-squared error (MSE) between the diffractive output quantitative phase images and their ground truth across all the wavelength channels, i.e., $\mathcal{L} = \frac{1}{N_w}\sum_{w=1}^{N_w} MSE(\boldsymbol{O}_w^{(GT)}, \boldsymbol{O}_w)$. Here $\boldsymbol{O}_w$ represents the normalized optical intensity distribution at a target spectral band $\lambda_w$ within the output FOV of the diffractive imager, and $\boldsymbol{O}_w^{(GT)}$ denotes the ground truth counterpart of $\boldsymbol{O}_w$. Further details regarding the training can be found in the Method section.

As a proof-of-concept, we numerically demonstrate diffractive multispectral QPI processor designs that operate in the visible spectral range, i.e., $\lambda_{N_w} = 450$ nm and $\lambda_1 = 700$ nm. We trained two multispectral QPI designs, which mainly differ in their number of target spectral bands $N_w$ and input spatial resolution ($N_x^{(\Psi)} \times N_y^{(\Psi)}$). The first design features $N_w = 9$ with $N_x^{(\Psi)} \times N_y^{(\Psi)} = 28 \times 28$, while the second design has $N_w = 16$ with $N_x^{(\Psi)} \times N_y^{(\Psi)} = 12 \times 12$ (see the Methods for details). Both of these diffractive designs are composed of 10 diffractive layers, where each diffractive layer has $1{,}000 \times 1{,}000$ and $784 \times 784$ diffractive features for the designs with $N_w = 9$ and 16, respectively. The entire diffractive volume spans an axial length of ~$112.7\lambda_m = 64.8$ μm and a lateral size of 225 μm (176.4 μm) for the diffractive design with $N_w = 9$ (16), forming a compact system that can be monolithically integrated with a CMOS image sensor. Without loss of generality, at the output plane of these diffractive designs, a monochrome image sensor with a pixel size of 1.8 μm × 1.8 μm (~$3.13\lambda_m \times 3.13\lambda_m$) is assumed to be used. A unit magnification is assumed between the object/input plane and the monochrome output/sensor plane, resulting in the same size of the output signal region $\mathcal{S}$ as the input FOV.

After the training process is complete, the resulting diffractive layer thickness profiles of the two designs are presented in **Figure 2a** and Figure S2a, Supporting Information. To assess the multispectral QPI performance of these diffractive designs, we used a test set of 5000 phase-only objects, wherein each object is created by randomly selecting images from the MNIST test set and encoding them into the multispectral object phase channels using a dynamic range of $[0, \alpha_{test}\pi]$. Two commonly used image quality metrics, SSIM and PSNR, were used to compare the resulting diffractive QPI measurements ($\boldsymbol{\Phi}$) and their ground truth counterparts ($\boldsymbol{\Psi}$) for all the target spectral bands. We summarized the resulting QPI performance under $\alpha_{test} = 1$ for the two diffractive processor designs with $N_w = 9$ and 16 in Figure 2b and Figure S2b, Supporting Information, respectively. The SSIM and PSNR values for the $N_w = 9$ design were calculated as $0.770 \pm 0.015$ and $17.04 \pm 0.33$ dB, respectively, while for the second design with $N_w = 16$ these same metrics were found as $0.726 \pm 0.031$ and $16.67 \pm 0.43$ dB, respectively.

We also visualized examples of these blind testing results in the bottom two rows of **Figure 3a** and Figure S3a, Supporting Information for the two diffractive processor designs with $N_w = 9$ and 16, respectively. The diffractive quantitative phase images clearly exhibit high structural fidelity matching their ground truths, further confirming the success of our designs. Additionally, we also present an image matrix that visualizes all the contributions from each channel of the input phase profile $\boldsymbol{\Psi}_w$ under illumination $\lambda_w$ to each of the output QPI measurement channels $\boldsymbol{\Phi}_s$. These individual contributions are denoted as $\boldsymbol{\phi}_{s,w}$, and can be expressed using the following form:

$$\boldsymbol{\phi}_{s,w} = \frac{\delta[|\boldsymbol{o}_w|^2, \mathcal{S}_s]}{Ref_s} \tag{5}$$

which denotes the intensity images measured by the array of monochrome sensor pixels corresponding to the channel $S_s$ using illumination at $\lambda_w$. Due to the spectral integration process occurring at the monochrome image sensor, as indicated in Equation 3, all the $\boldsymbol{\phi}_{s,w}$ values in the same column will add up to form $\boldsymbol{\Phi}_s$. Our results shown in Figure 3a and Figure S3a, Supporting Information demonstrate that the input phase profiles of each spectral band are successfully converted into the correct intensity distributions within their corresponding output QPI images (i.e., the diagonal entries with $s = w$); there are also some weak cross-talk terms corresponding to the other spectral bands (i.e., the off-diagonal entries with $s \neq w$).

To better quantify the impact of this spectral cross-talk on our multispectral QPI results, we calculated the mean optical power of $\boldsymbol{\phi}_{s,w}$ using the diffractive design with $N_w = 9$ over the entire testing set, and summarized the results into the confusion matrices reported in Figure 3b and c. A similar cross-talk analysis for the other diffractive QPI design with $N_w = 16$ was also reported by the confusion matrices presented in Figure S3b and c, Supporting Information. In all of these analyses and the corresponding confusion matrices, each row represents a distinct illumination wavelength ($\lambda_w$), while the columns correspond to individual sensor pixel groups ($S_s$), each designated to a particular spectral band $\lambda_s$. In the confusion matrices presented in Figure 3b and Figure S3b, Supporting Information, the optical power values, presented as percentages, are normalized using the total power of a specific illumination wavelength ($\lambda_*$) received by the entire sensor signal region ($\mathcal{S}$), resulting in a summation of 100% in each row. Based on these analyses, we found that, for the two diffractive designs with $N_w = 9$ and 16, when a given illumination $\lambda_w$ is used, the power received by its corresponding sensor pixel group designated to this wavelength (corresponding to the diagonal entries $\boldsymbol{\phi}_{s,w}|_{s=w}$) was, on average, $(7.38 \pm 1.24)$ and $(9.95 \pm 1.93)$-fold greater, respectively, than the mean power received by the other sensor pixel groups (corresponding to spectral power leakage, $\boldsymbol{\phi}_{s,w}|_{s \neq w}$). Therefore, the amount of energy directed to incorrect sensor pixel groups is minimal (as desired) compared to that directed to the intended group of pixels. Similarly, the confusion matrices in Figure 3c and Figure S3c, Supporting Information present another view on the magnitudes of these cross-talk components. Here, the normalization is performed using the total power received by a particular sensor pixel group ($\mathcal{S}_*$), leading to a column-wise summation of 100%. From this perspective, given a group of sensor pixels $\mathcal{S}_s$, the power received at its designated target spectral band (corresponding to $\boldsymbol{\phi}_{s,w}|_{s=w}$) is found to be on average $(7.32 \pm 0.19)$ and $(9.92 \pm 0.38)$-fold larger, respectively, than the mean power of the other spectral bands (corresponding to spectral power leakage, $\boldsymbol{\phi}_{s,w}|_{s \neq w}$) for the two designs with $N_w = 9$ and 16, indicating that these spectral-spatial cross-talk components produced by our diffractive QPI design also minimally affect its signal contrast. We also created an alternate version of these confusion matrices resulting from a global normalization, using the total optical power received across the entire sensor signal region ($\mathcal{S}$) for all the spectral bands, which is provided in Figure S4, Supporting Information, further supporting the same conclusions.

We should also note that the QPI signal output power at different wavelength channels for both of the diffractive imager designs demonstrates a negative correlation with the illumination wavelength, as shown in Figure 3b and Figure S3b, Supporting Information. Stated differently, our diffractive QPI processors tend to direct smaller wavelengths to their corresponding virtual spectral filter locations more efficiently than larger wavelengths; for example, the sensor pixel group assigned to the shortest wavelength channel $\lambda_{N_w}$ exhibits the strongest response. In addition, Figure 2b and Figure S2b, Supporting Information reveal that the QPI performance of the individual spectral channels in terms of PSNR and SSIM values also improves at shorter wavelengths. These observations can be ascribed to the diffraction limit of light: compared to longer wavelengths, the effective number of trainable diffractive features that shorter wavelengths can control is larger, resulting in richer degrees of freedom to obtain better diffraction efficiencies and QPI performance at shorter wavelengths.

In the analyses presented in Figure 2 and 3, we used a constant $\alpha_{test}$ of 1, meaning that the input phase in each wavelength channel has a maximum phase contrast of π. To more comprehensively test the QPI performance of our diffractive imager designs, we further tested them using a different phase contrast factor ($\alpha_{test}$) than the one used in the training ($\alpha_{tr,max} = 1$). A total of 9 different values were selected within a range of [0.2, 1.8] as the values of $\alpha_{test}$, and the same MNIST test image dataset is used for the phase encoding of the input objects. The resulting multispectral QPI images using the 9 and 16-channel diffractive designs are shown in **Figure 4b** and Figure S5b, Supporting Information, respectively, and their reconstruction quality was also quantified using the same SSIM and PSNR metrics (reported in Figure 4a and Figure S5a, Supporting Information) across all the target spectral bands and test objects. These analyses reveal that both the SSIM and PSNR values peak at $\alpha_{test} = 1$, which matches the maximum phase contrast of the objects used during the training stage ($\alpha_{tr} \in U[0.1,1]$). To the left of these peaks, where $\alpha_{test} < 1$, the multispectral QPI performance of our diffractive processors appears to degrade slightly. For instance, compared to their peak values obtained at $\alpha_{test} = 1$, the resulting SSIM and PSNR values at $\alpha_{test} = 0.2$ show a reduction of ~7.5% and ~12.8%, respectively, which can primarily be ascribed to the increased demand for phase sensitivity when resolving a 5-fold smaller input phase contrast[59]. However, when it comes to the case where the testing input phase contrast exceeds π, i.e., $\alpha_{test} > \alpha_{tr,max} = 1$, the performance of our diffractive QPI designs exhibits more degradation. This becomes particularly noticeable as $\alpha_{test}$ approaches 2, which is expected as the training process did *not* involve any input objects with a phase contrast of $\alpha_{test} > 1$, thus limiting the diffractive processor's ability to generalize to this range.

To gain deeper insights into the diffractive multispectral QPI processor's ability to resolve input object information with lower phase contrast, we performed an additional analysis by investigating the impact of the feature resolution and phase contrast of the input objects on the multispectral QPI performance of the trained diffractive networks. To standardize our tests, for the 9-channel multispectral QPI design, we created binary phase grating patterns with linewidths of $18.4\lambda_m =$ ~10.8 μm and $9.2\lambda_m =$ ~5.4 μm, and selected the test phase contrast parameter $\alpha_{test}$ from {0.05, 0.1, 0.2}. The diffractive QPI signals resulting from these test objects using the 9-channel diffractive imager design shown in Figure 2a are depicted in **Figure 5**. Our results demonstrate that this diffractive imager with $N_w = 9$ can successfully resolve the test phase gratings with a linewidth of $18.4\lambda_m$ for $\alpha_{test} \geq 0.1$. When using $\alpha_{test} = 0.1$, if the test phase object is changed to the one with a 2-fold narrower linewidth of $9.2\lambda_m$, the diffractive QPI outputs become slightly harder to resolve for a small portion of the target spectral bands (e.g., $\lambda_1$ and $\lambda_9$). When using an input phase object with $\alpha_{\text{test}} > 0.1$, our diffractive QPI design can clearly resolve spatial features with a linewidth of $\geq 9.2\lambda_{\text{m}} = \sim 5.4$ µm at all the 9 spectral bands of interest. Similarly, for the other diffractive QPI design with $N_w = 16$ target spectral bands, we also performed a spatial resolution and phase sensitivity analysis using binary phase patterns with linewidths of $25\lambda_{\text{m}} = \sim 14.4$ µm and $12.5\lambda_{\text{m}} = \sim 7.2$ µm. Figure S6, Supporting Information reveals that when using an input phase object with $\alpha_{\text{test}} \geq 0.1$ our diffractive QPI design can resolve linewidths of $\geq 12.5\lambda_{\text{m}} = \sim 7.2$ µm at all the 16 spectral bands of interest.

These diffractive multispectral QPI processor designs reported so far were trained using a dataset of relatively simple spatial patterns, including handwritten digits and grating-like patterns. To further evaluate the generalization performance of our diffractive QPI designs on inputs with distinct spatial distributions, we conducted additional numerical tests using human Pap smear microscopy images, which exhibit substantially different spatial features compared to the training image datasets. These blinded test results are showcased in **Figure 6a**, revealing a decent agreement between the diffractive multispectral QPI results and the corresponding ground truth images. We calculated the image quality metrics across the entire Pap smear test dataset (see Figure 6b), where the SSIM and PSNR values were 0.863 ± 0.039 and 20.37 ± 2.20 dB, respectively. Overall, these successful external generalization test results demonstrate that our diffractive multispectral QPI design is not limited to specific object types or features, and can broadly serve as a multispectral quantitative phase imager for various objects.

## 3 DISCUSSION

The diffractive multispectral QPI framework reported in this work exhibits certain limitations. First, due to the wavelength-dependent diffraction effects, the QPI performance across the different spectral channels presents variations, with a maximum deviation of ~6% and ~8% observed in the 9- and 16-channel diffractive QPI designs, respectively. To address this issue, one can use a weight coefficient for the loss terms of the different spectral bands, which can be adaptively updated (increased/decreased dynamically) during the training stage based on the QPI performance imbalance among the spectral channels[62]. Second, as an initial proof-of-concept, our diffractive QPI designs presented in this manuscript were trained without optimization of the output diffraction efficiencies, which led to relatively low average diffraction efficiency values of ~0.79% and ~0.26% for the 9- and 16-channel diffractive QPI designs, respectively. A viable solution to increase the output diffraction efficiency of a diffractive QPI processor could involve incorporating an additional penalty term related to diffraction efficiency into the training loss function, albeit at the cost of a slight degradation in QPI performance. Third, during the actual implementation of our diffractive QPI designs, the performance of our diffractive networks can be markedly affected by misalignments and manufacturing errors. This issue can be partially mitigated by modeling these imperfections and incorporating them into our physical forward model in the form of random variations, which is referred to as the "vaccination" of diffractive networks[55]; vaccinated diffractive optical networks are in general more resilient against fabrication imperfections and exhibit a significantly better match between their numerical and experimental results[55,58,60,61,63,67,68].

In our diffractive multispectral QPI designs, each voxel in the desired multispectral data cube maps to a pixel in the output signal region ($\mathcal{S}$) of the 2D monochrome sensor array; therefore, the total pixel count within $\mathcal{S}$ will be equal to the product of the number of wavelength channels ($N_w$) and the number of pixels per channel ($N_o$). The total pixel count of an image sensor array or focal plane array can be limited and hard to increase, such as in the case of infrared and terahertz parts of the spectrum. Thus, a careful balance must be considered in selecting $N_w$ and $N_o$ according to the needs of an application and the availability of high-pixel count image sensor arrays. Furthermore, to fully leverage the pixel count (or sensing throughput) of a monochrome image sensor array for a QPI processor design, the diffractive optical front-end must achieve imaging with an adequate space bandwidth product (SBP). To achieve QPI with a high SBP, or to perform high-resolution QPI with a larger number of spectral channels, it would be prudent to enhance the depth of the diffractive optical network and concurrently increase the total number of diffractive neurons $N$, so that the degrees of freedom provided by the diffractive processor can be enhanced to match the spatial and spectral complexities of the target transformation[56,57,60,62,69,70]. Also, a deeper diffractive architecture (i.e., with a larger number of diffractive layers, one following another) generally provides strong benefits in terms of output power efficiency, transformation accuracy, and spectral multiplexing capability[57,64,65].

In summary, we demonstrated the design of diffractive multispectral QPI processors to simultaneously measure the input object phase information, across a large set of target spectral bands, without the need for any specific color filter arrays and computational algorithms. These diffractive multispectral QPI processors maintain their performance and phase accuracy despite variations in the intensity of the broadband light sources used for illumination. With the selection of appropriate nano-/micro-fabrication methods, such as two-photon polymerization-based 3D printing[68,71,72], these diffractive optical processors can be physically scaled (expanded/shrunk) to function within different parts of the electromagnetic spectrum.

## 4 METHODS

**Optical forward model of the diffractive multispectral QPI processors**

Our multispectral QPI framework is composed of $K$ successively positioned diffractive layers, each containing thousands of spatially coded diffractive features. In our numerical forward model, these diffractive layers are treated as thin planar elements that introduce complex-valued transmission modulation to the incident coherent optical fields. For the $q^{th}$ diffractive feature located on the $l^{th}$ layer at the spatial position $(x_q, y_q, z_l)$, its complex-valued transmission coefficient can be represented as a function of the material thickness value $h_q^l$, which is given by:

$$t(x_q, y_q, z_l; \lambda) = \exp\left(\frac{-2\pi\kappa(\lambda)h_q^l}{\lambda}\right) \exp\left(\frac{-j2\pi(n(\lambda) - n_{\text{air}})h_q^l}{\lambda}\right) \quad (6).$$

Here, $n(\lambda)$ and $\kappa(\lambda)$ represent the refractive index and the extinction coefficient of the used dielectric material, respectively, which correspond to the real and imaginary parts of the complex-valued refractive index $\tilde{n}(\lambda)$, i.e., $\tilde{n}(\lambda) = n(\lambda) + j\kappa(\lambda)$. In the numerical forward model of all the diffractive processor designs reported in this paper, we selected BK7 as the material of the diffractive layers[73], with the refractive index curve $n(\lambda)$ provided in Figure S7, Supporting

Information. Since the absorption of this material within the visible spectrum is negligible, $\kappa(\lambda)$ is set to 0. The thickness value $h$ of each diffractive feature is defined as a summation of two parts $h_{\text{learnable}}$ and $h_{\text{base}}$, namely:

$$h = h_{\text{learnable}} + h_{\text{base}} \tag{7}$$

where $h_{\text{learnable}}$ refers to the learnable thickness value of each diffractive feature and is constrained within the range $[0, h_{\max}]$, and $h_{\text{base}}$ is a constant representing the additional base thickness that acts as the substrate support for the diffractive features. In the diffractive designs of this paper, $h_{\max}$ is set as 664.5 nm, corresponding to a full phase modulation range from 0 to $2\pi$ for the largest wavelength ($\lambda_1$). $h_{\text{base}}$ is empirically selected as 700 nm.

To numerically model the behavior of the free-space propagation of light between the diffractive layers, we adopted the Rayleigh-Sommerfeld scalar diffraction theory, where the diffraction process can be formulated as a linear, shift-invariant operator with an impulse response. We defined the $q^{\text{th}}$ diffractive feature on the $l^{\text{th}}$ layer at $(x_q, y_q, z_l)$ as the source of a secondary wave, generating a complex field at $\lambda$ given by the following equation:

$$w_q^l(x, y, z; \lambda) = \frac{z - z_l}{(r_q^l)^2} \left( \frac{1}{2\pi r_q^l} + \frac{n}{j\lambda} \right) \exp\left( \frac{j 2\pi n r_q^l}{\lambda} \right) \tag{8}$$

where $r_q^l = \sqrt{(x - x_q)^2 + (y - y_q)^2 + (z - z_l)^2}$. These secondary waves created by the diffractive features on the $l^{\text{th}}$ layer propagate to the next layer (the $(l+1)^{\text{th}}$ layer) and are spatially superimposed. Therefore, the optical field incident on the $p^{\text{th}}$ diffractive feature of the $(l+1)^{\text{th}}$ layer at $(x_p, y_p, z_{l+1})$ can be formulated as the convolution of the complex amplitude $u_q^l$ of the wave field right after the $q^{\text{th}}$ diffractive feature on the $l^{\text{th}}$ layer with the impulse response function $w_q^l(x_p, y_p, z_{l+1}; \lambda)$. This resulting field is then modulated by the transmittance $t(x_p, y_p, z_{l+1}; \lambda)$ of the $(l+1)^{\text{th}}$ diffractive layer, which can be expressed as:

$$u_p^{l+1}(x, y, z; \lambda) = t(x_p, y_p, z_p; \lambda) \sum_q u_q^l w_q^l(x_p, y_p, z_{l+1}; \lambda) \tag{9}$$

In the numerical modeling of all the diffractive designs in this paper, the spatial sampling rate of the simulated complex fields is set to be half of the shortest wavelength, i.e., $0.5\lambda_{N_w} = 225$ nm, which also corresponds to the lateral size of a diffractive feature. All the axial distances between the adjacent layers (including the diffractive layers as well as the input/output planes) are set to ~$9.39\lambda_m$.

**Numerical implementation of the diffractive multispectral QPI processors**

A dispersive phase-only object is set to be located at $z = z_0$, which has a uniform distribution of unit amplitude across all the spectral bands of interest and phase profiles $\Psi_w(x, y)$ that depend on the illumination wavelength $\lambda_w$, where $\boldsymbol{\Psi}_w \in \mathbb{R}^{N_x^{(\Psi)} \times N_y^{(\Psi)}}$. This object is illuminated by a broadband, spatially coherent source, resulting in multispectral input complex fields that can be written in the following form:

$$i_w(x,y) = e^{j\Psi_w(x,y)} \quad (10),$$

where $i_w$ represents the input complex field at $\lambda_w$. $i$ can also be represented using $u^1$, i.e., $u^1(x,y,z_0;\lambda_w) = e^{j\Psi_w(x,y)}$. From this input field $\boldsymbol{i}$ (or $\boldsymbol{u^1}$), the processes of diffractive layer modulation and secondary wave generation outlined in the last subsection are successively performed as the complex fields propagate through the diffractive network volume. This continues until the output plane of the diffractive processor, ultimately forming an output complex field $o_w(x,y) = u^K(x,y,z_K;\lambda_w)$. A monochrome image sensor array measures the output intensity distribution $\boldsymbol{D}$ within this output FOV. Without loss of generality, we assumed an ideal, flat spectral responsivity of the image sensor, and formulated $\boldsymbol{D}$ as the integration of the diffractive output intensities across different wavelength channels. The expression of $\boldsymbol{D}$ can be found in Equation 1.

According to the layout presented in Figure S1, Supporting Information, the pixels in $\boldsymbol{D}$ are spatially divided into two regions: an output signal region $\mathcal{S}$ and a reference signal region $\mathcal{R}$, which are used to calculate the normalized output signal that represents the desired quantitative phase information. The detailed spatial layout of these signal subregions is also illustrated in Figure S1, Supporting Information. The output signal region $\mathcal{S}$ has a pixel number of $N_x^{(\mathcal{S})} \times N_y^{(\mathcal{S})} = (N_x^{(D)} - 2) \times (N_y^{(D)} - 2)$, which can be divided into $N_x^{(\Psi)} \times N_y^{(\Psi)}$ blocks, each corresponding to an individual spatial pixel of the input object. Moreover, each one of these blocks consists of $\sqrt{N_w} \times \sqrt{N_w}$ discrete output pixels, where each output pixel is uniquely assigned to each one of the $N_w$ wavelength channels; therefore we have $N_x^{(\Psi)} = (N_x^{(D)} - 2) / \sqrt{N_w}$ and $N_x^{(\Psi)} = (N_y^{(D)} - 2) / \sqrt{N_w}$. To obtain the quantitative phase information that corresponds to a certain wavelength channel $\lambda_s$, a spatial demosaicing process $\delta$ is performed to the output intensity values within the output signal region $\mathcal{S}$, which can be mathematically expressed as:

$$\delta[\boldsymbol{D},\mathcal{S}_s](m,n) = D_{\mathcal{S}}\left((m-1)\sqrt{N_w} + \left\lfloor\frac{s-1}{\sqrt{N_w}}\right\rfloor + 1, (n-1)\sqrt{N_w} + s - \left\lfloor\frac{s-1}{\sqrt{N_w}}\right\rfloor\sqrt{N_w}\right) \quad (11),$$

where $D_{\mathcal{S}}$ represents the part of $D$ located in the region $\mathcal{S}$, $\mathcal{S}_s$ denotes the collection of all the pixels within $\mathcal{S}$ that are assigned to $\lambda_s$, and $\lfloor \cdot \rfloor$ denotes the floor operation. With this demosaicing operation, the pixels belonging to $\mathcal{S}_s$ can be extracted to form a new image with a size of $N_x^{(\Psi)} \times N_y^{(\Psi)}$. This is followed by the normalization operation using the calculated multispectral reference signal $Ref_s$; see Equation 3.

For the diffractive multispectral QPI processor design shown in Figure 2a that operates with $N_w = 9$, the size of the input/output FOV is set to be ~$262.96\lambda_m \times 262.96\lambda_m$, indicating a unit magnification system between the object and sensor planes. The input object has a pixel number of $N_x^{(\Psi)} \times N_y^{(\Psi)} = 28 \times 28$ which is the same as the default resolution of the MNIST images, resulting in each input pixel having a size of ~$9.39\lambda_m \times 9.39\lambda_m$. The output signal region $\mathcal{S}$ consists of $N_x^{(\mathcal{S})} \times N_y^{(\mathcal{S})} = 84 \times 84$ pixels, resulting in each output pixel having a size of ~$3.13\lambda_m \times 3.13\lambda_m$. In the demosaicing process, these 84×84 pixels are grouped into 28×28 blocks that individually correspond to the spatial pixels of the input object, where each block contains 3×3 pixels, corresponding to the $N_w = 9$ target spectral bands. To achieve the multispectral QPI task, we designed the diffractive multispectral QPI processor to possess 1,000 × 1,000 diffractive features per layer, resulting, in each layer, a diffractive area of ~$390\lambda_m \times 390\lambda_m$.

For the diffractive multispectral QPI processor design shown in Figure 2a, Supporting Information with $N_w = 16$ spectral bands, the size of the input/output FOV is set to be ~$153.39\lambda_m \times 153.39\lambda_m$. The input object has a pixel number of $N_x^{(\Psi)} \times N_y^{(\Psi)} = 12 \times 12$, which is slightly smaller than the default resolution of the MNIST images, resulting in each input pixel having a size of ~$12.52\lambda_m \times 12.52\lambda_m$. The output signal region $\mathcal{S}$ consists of $N_x^{(\mathcal{S})} \times N_y^{(\mathcal{S})} = 48 \times 48$ pixels. In the demosaicing process, these 48×48 pixels are grouped into 12×12 blocks that individually correspond to the spatial pixels of the input object, where each block contains 4×4 pixels, corresponding to the $N_w = 16$ target spectral bands. The diffractive processor is designed to possess 784 × 784 diffractive features per layer, resulting, in each layer, a diffractive area of ~$307\lambda_m \times 307\lambda_m$.

**Training loss function and image quality metrics**

To train our diffractive QPI processors, we devised a normalized MSE-based loss function to penalize the structural difference between the diffractive output intensity distribution $O_w(x,y)$ and its ground truth counterpart $O_w^{(\text{GT})}(x,y)$ at each spectral channel $\lambda_w$. The loss for a specific wavelength channel was defined as:

$$\mathcal{L}_w = \frac{1}{N_x^{(D)}} \frac{1}{N_y^{(D)}} \sum_{x=1}^{N_x^{(D)}} \sum_{y=1}^{N_y^{(D)}} \left| \sigma_w^{(\text{GT})} O_w^{(\text{GT})}(x,y) - \sigma_w O_w(x,y) \right|^2 \quad (12),$$

Here $\sigma_w^{(\text{GT})}$ and $\sigma_w$ represent the normalization factors used to normalize the energy of the output image $\boldsymbol{O}_w$ from the diffractive processor with regard to $\boldsymbol{O}_w^{(\text{GT})}$, i.e.,

$$\sigma_w^{(\text{GT})} = \frac{1}{\sum_{x=1}^{N_x^{(D)}} \sum_{y=1}^{N_y^{(D)}} O_w^{(\text{GT})}(x,y)} \quad (13),$$

$$\sigma_w = \frac{\sum_{x=1}^{N_x^{(D)}} \sum_{y=1}^{N_y^{(D)}} \sigma_w^{(\text{GT})} O_w^{(\text{GT})}(x,y) O_w(x,y)}{\sum_{x=1}^{N_x^{(D)}} \sum_{y=1}^{N_y^{(D)}} O_w^2(x,y)} \quad (14).$$

In the training stage, the loss function used for the multispectral QPI processor is calculated by averaging the loss terms across all the $N_w$ spectral channels, so that all the channels were optimized simultaneously. Therefore, the total loss function can be written as:

$$\mathcal{L} = \frac{1}{N_w} \sum_{w=1}^{N_w} \mathcal{L}_w \qquad (15),$$

The structural fidelity of the quantitative phase images $\Phi_s(x, y)$ produced by the diffractive multispectral QPI processor against their ground truth $\Psi_w(x, y)$ was evaluated using the SSIM and PSNR metrics. For a specific wavelength channel $\lambda_w$, the PSNR value is quantified using the following equation:

$$PSNR_{s \equiv w} = 20\log_{10}\left(\frac{1}{\sqrt{\sum_x \sum_y |\Psi_w(x,y) - \Phi_s(x,y)|^2}}\right) \qquad (16).$$

The calculation of the SSIM metric was performed by using the TensorFlow built-in function *tf.image.ssim()*, which implements the standard definition of SSIM[74] with default parameters.

**Training data preparation and other implementation details**

For the training of our diffractive multispectral QPI processors, we formed a training image dataset composed of 110,000 images, with two distinct parts: (1) a set of 55,000 handwritten digit images extracted from the original MNIST training dataset, and (2) a set of 55,000 custom-created images of various patterns. These patterns[66] include different shapes, such as gratings, patches and circles, each exhibiting unique spatial periods and orientations. During the training process, we created each input object by randomly selecting $N_w$ images from 110,000 images in the training set and encoding each of these selected images into one of the object phase channels at the $N_w$ predetermined wavelengths, forming dispersive phase objects.

The numerical simulations and training of the diffractive multispectral QPI processor designs presented in this work were implemented using TensorFlow (version 2.5.0). We used the Adam optimizer with the default parameters in TensorFlow. The learning rate and batch size were set to 0.001 and 16, respectively. The diffractive models were trained for 100 epochs using a workstation with an Nvidia GeForce RTX 3090 GPU, an Intel Core i9-11900 CPU and 128 GB of RAM. The training of the 10-layer diffractive multispectral QPI processor design shown in Figure 2 took approximately 12 days. This training process of our diffractive multispectral QPI processor, despite taking a relatively long time, is a one-time design effort.

**Supporting Information:** This file contains Figures S1-S7, Supporting Information.

# Figures

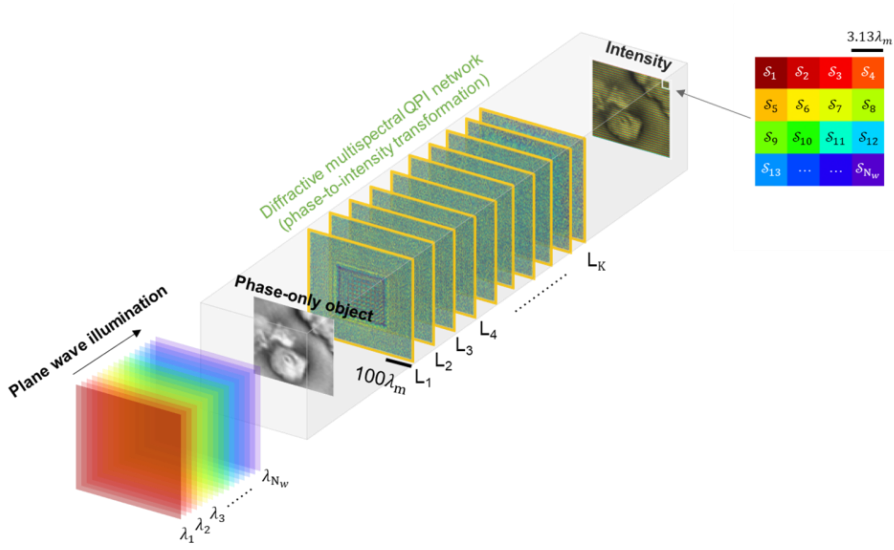

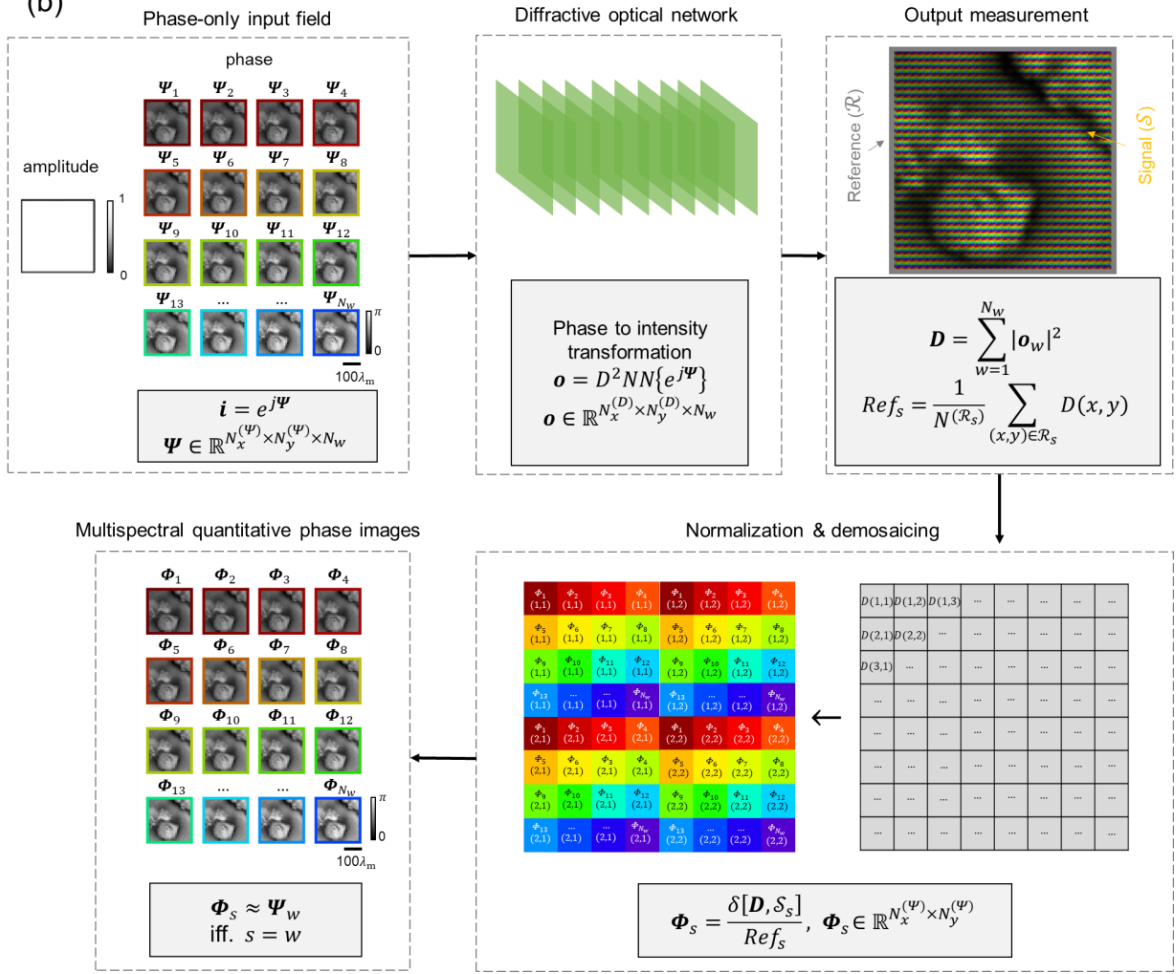

**Figure 1. Schematic and working principle of multispectral quantitative phase imaging (QPI) using a diffractive processor. a,** Optical layout of the diffractive processor. This diffractive

processor is composed of *K* diffractive layers, which are jointly trained using deep learning to simultaneously perform phase-to-intensity transformations at a predetermined set of spectral bands $\{\lambda_1, \lambda_2, …, \lambda_{N_w}\}$, while also routing the resulting multispectral QPI signals to different, desired spatial positions within the output FOV. **b,** Pipeline of the presented multispectral QPI framework. A monochrome image sensor array is employed at the output plane to measure the resulting intensity distribution, where the pixels within the output signal region ($\mathcal{S}$) are designated to $N_w$ subgroups, each carrying the QPI signal corresponding to a unique wavelength channel. These QPI signals are then normalized using the corresponding reference signals collected within the reference signal region ($\mathcal{R}$) surrounding $\mathcal{S}$, resulting in the output quantitative phase images $\boldsymbol{\Phi}$ that approximate the ground truth phase profiles $\boldsymbol{\Psi}$ at the target spectral bands $\{\lambda_1, \lambda_2, …, \lambda_{N_w}\}$.

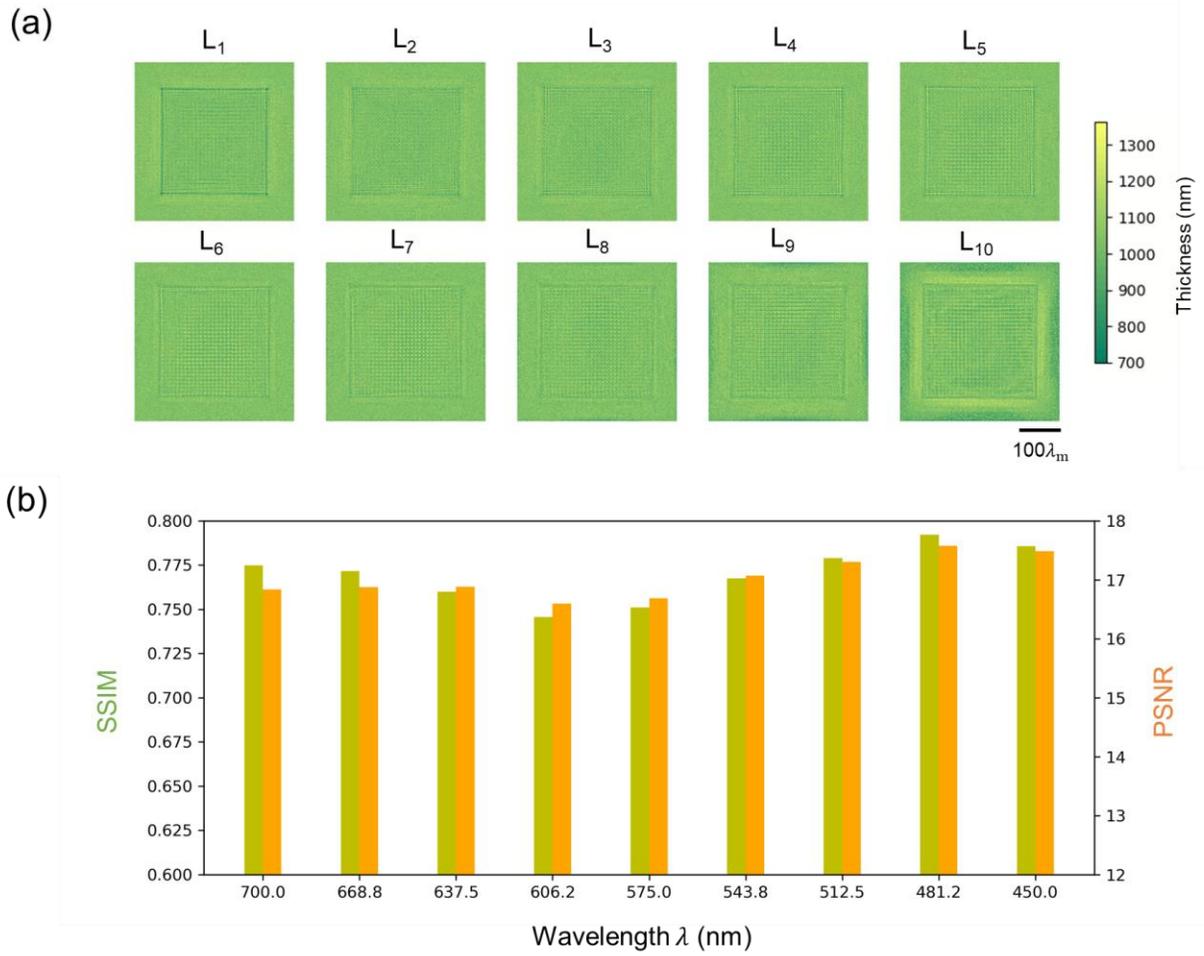

**Figure 2. The diffractive multispectral QPI processor design with $N_w = 9$ target spectral bands. a,** Thickness profiles of the trained diffractive layers. **b,** SSIM and PSNR values of the diffractive QPI processor outputs at different target spectral bands.

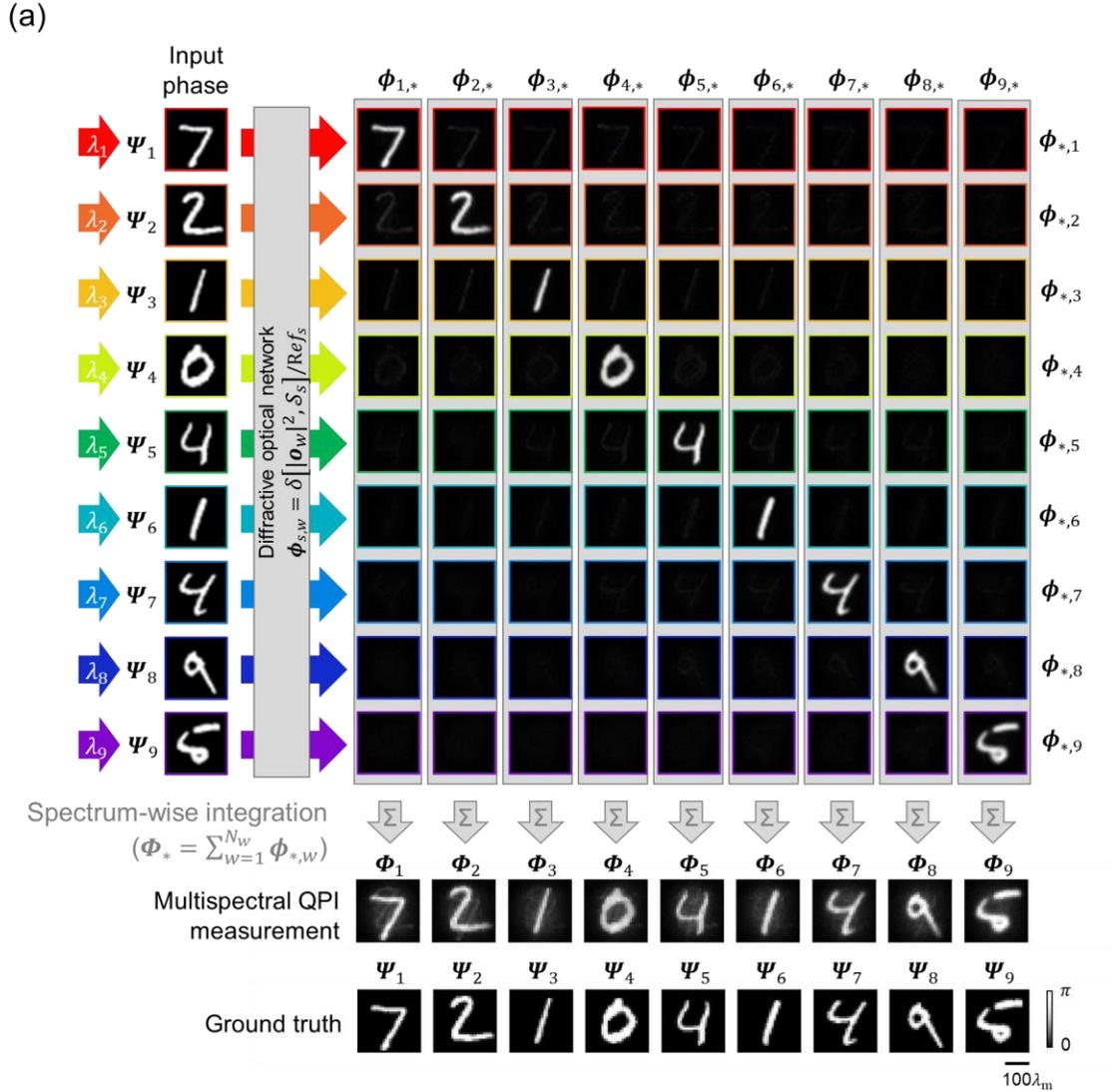

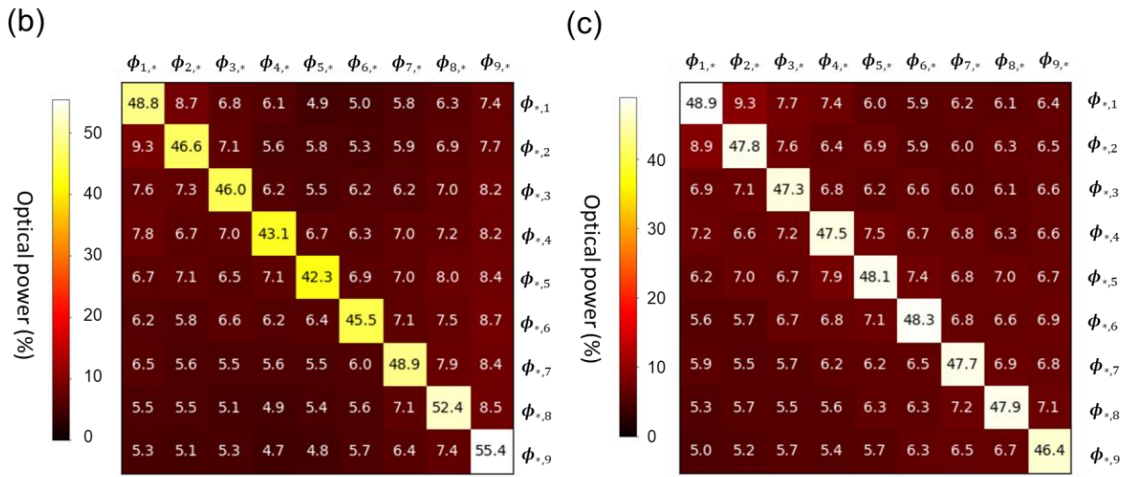

**Figure 3. Spectral cross-talk analysis for the outputs of the $N_w = 9$ channel diffractive multispectral QPI processor design shown in Figure 2a.** Output image matrix illustrating the spectral cross-talk, represented by the off-diagonal images. By summing up all the images in each column, the output measurements of the diffractive multispectral QPI processor, $\boldsymbol{\Phi}$, are obtained and visualized at the bottom of the image matrix as a separate row, which are compared to their corresponding ground truth $\boldsymbol{\Psi}$. **b,** Confusion matrix showing the distribution of the optical power received by different groups of sensor pixels $\mathcal{S}_s$ as a function of the illumination wavelength $\lambda$. Within this matrix, all the optical power percentages in each row add up to 100%, and the off-diagonal entries indicate the level of spectral cross-talk between different spectral bands. **c,** Same as (b), except that all the optical power percentages in each column add up to 100%.

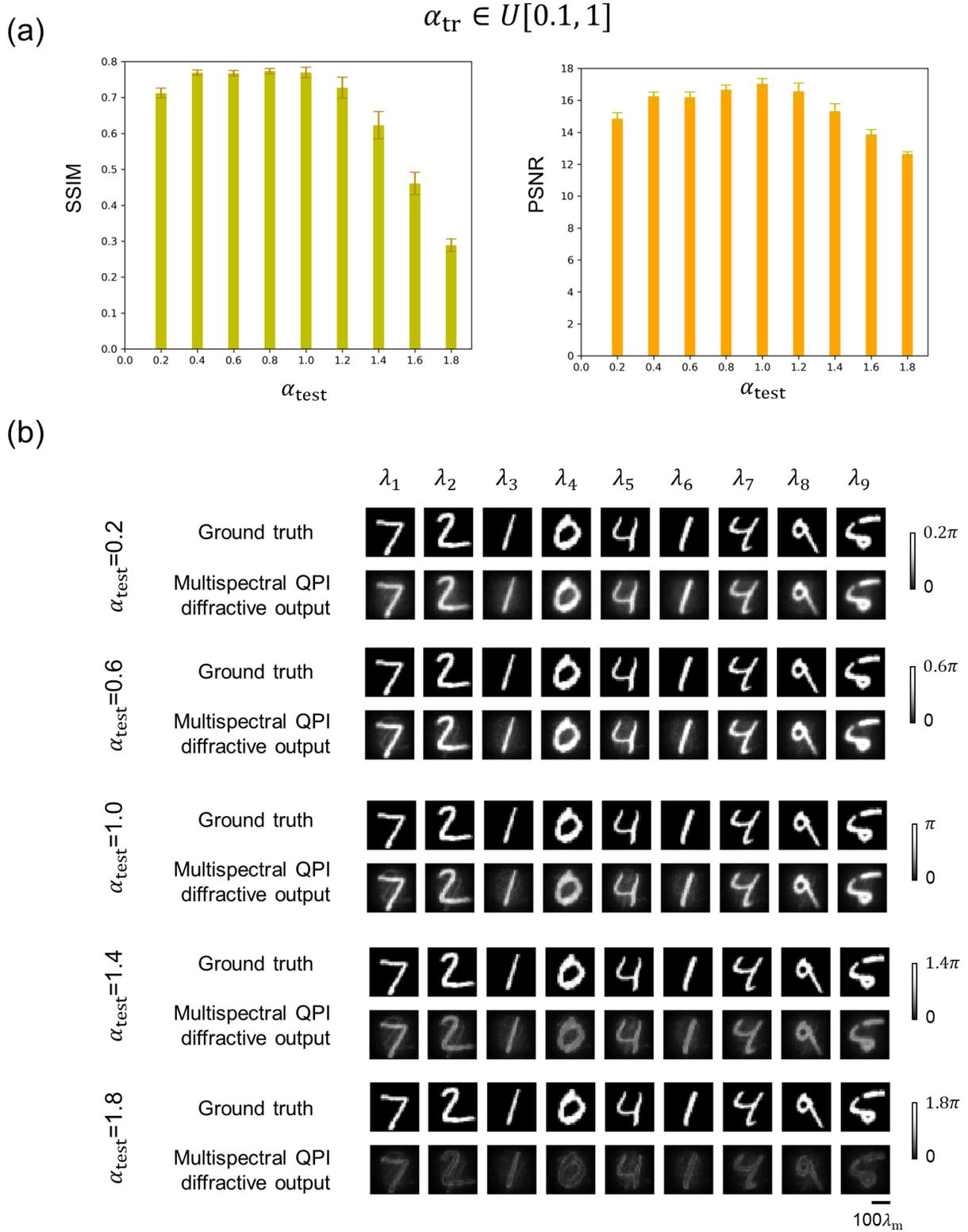

**Figure 4. Analysis of the impact of the input object phase contrast on the image quality of the diffractive multispectral QPI results using the $N_w = 9$ channel design shown in Figure 2a.** SSIM and PSNR values of the resulting QPI measurements ($\Phi$) as a function of the phase

contrast parameter $\alpha_{\text{test}}$. **b,** Examples of the multispectral QPI results using different $\alpha_{\text{test}}$ for the input test objects, never seen before during the training.

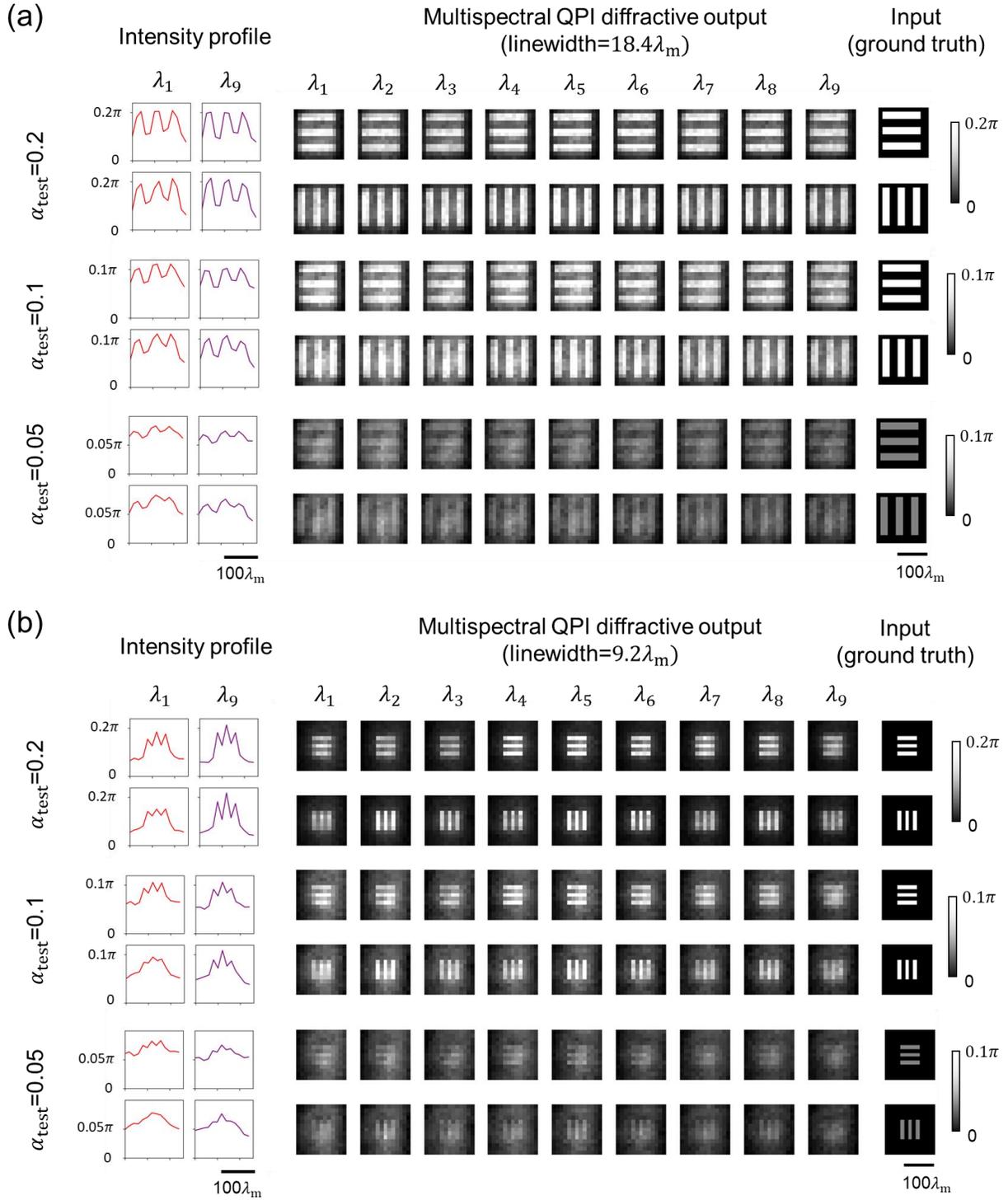

**Figure 5. Spatial resolution and phase sensitivity analysis for the $N_w = 9$ channel diffractive multispectral QPI processor design shown in Figure 2a.** Images of the binary phase grating patterns encoded within the phase channels of the input object, along with the resulting diffractive output QPI signals ($\Phi$) at the target spectral bands. The grating has a linewidth of $18.4\lambda_m$, and the phase contrast parameter ($\alpha_{test}$) of the input phase object is selected from {0.05, 0.1, 0.2}. The intensity profiles are calculated based on averaging across the grating area along the direction perpendicular to the gratings. **b,** Same as (a) except that the linewidth of the grating phase pattern is selected as $9.2\lambda_\text{m}$. Certain pixels in the diffractive QPI output with $\alpha_\text{test} = 0.05$ exhibit intensity values substantially higher than $0.05\pi$; therefore, for clarity purposes, the color bars used for these results and their ground truth are selected to have a larger range of $[0, 0.1\pi]$.

(a)

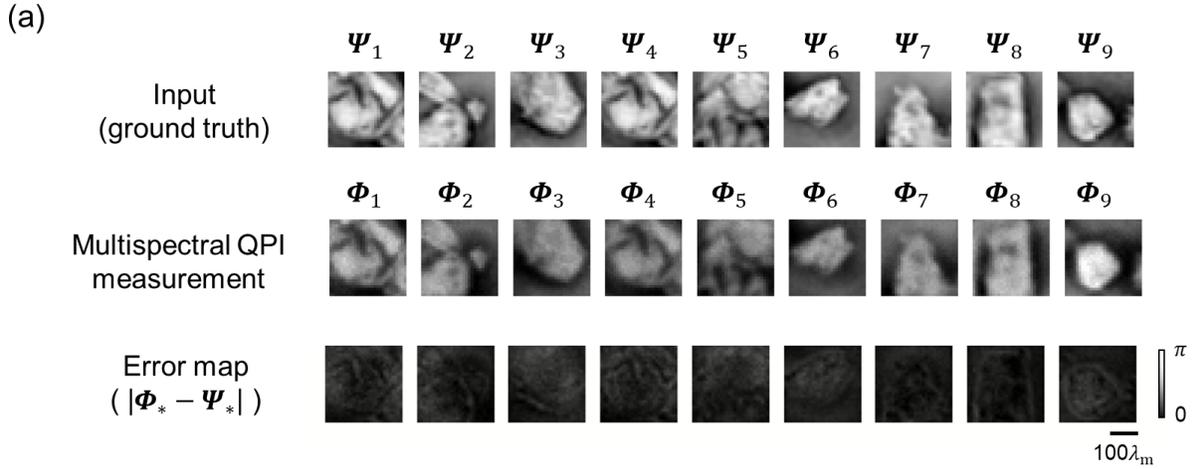

(b)

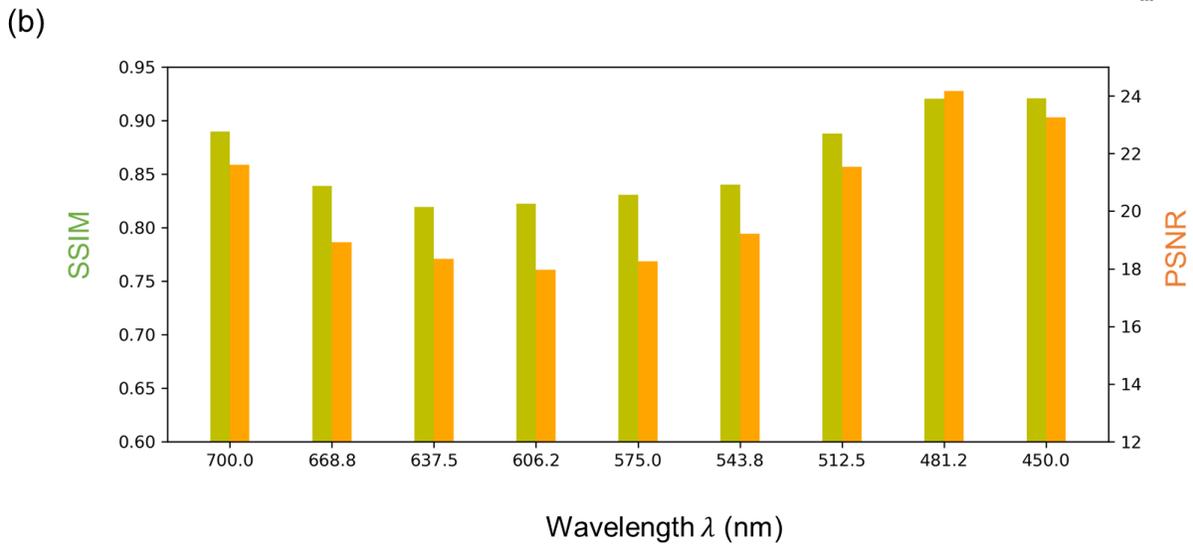

**Figure 6. Results for testing the external generalization performance of the $N_w = 9$ channel diffractive multispectral QPI processor design using blind testing images from a new dataset composed of Pap Smear images. a,** Examples of the input (ground truth) phase images at different spectral bands, which are compared to their corresponding diffractive QPI measurements. $\alpha_{\text{test}} = 1$ was used in the testing. **b,** SSIM and PSNR values of the diffractive multispectral QPI processor outputs at different target spectral bands.

29